\documentclass[pre,showpacs,nofootinbib,twocolumn,showkeys]{revtex4-1}
\usepackage{graphicx}
\usepackage{amsfonts}
\usepackage{amssymb}
\usepackage{amsmath}
\usepackage{array}
\usepackage{dcolumn}
\usepackage{color}
\usepackage{verbatim}
\usepackage{siunitx}
\usepackage{subcaption}

\begin{document}
\title{Fabrication of High Aspect Ratio Micro-Penning-Malmberg Gold Plated Silicon Trap Arrays}
\author{Alireza Narimannezhad}
\email{a.narimannezhad@wsu.edu}
\author{Joshah Jennings}
\author{Marc H. Weber}
\author{Kelvin G. Lynn}
\email{kgl@wsu.edu}
\affiliation{Center for Materials Research, Washington State University, Pullman, WA 99164-2711}

\begin{abstract}
{Acquiring a portable high density charged particles trap might consist of an array of micro-Penning-Malmberg traps (microtraps) with substantially lower end barriers potential than conventional Penning-Malmberg traps \cite{Simulation}. We report on the progress of the fabrication of these microtraps designed for antimatter storage such as positrons. The fabrication of large length to radius aspect ratio ($1000:1$) microtrap arrays involved advanced techniques including photolithography, deep reactive ion etching (DRIE) of silicon wafers to achieve through-vias, gold sputtering of the wafers on the surfaces and inside the vias, and thermal compression bonding of the wafers. This paper describes the encountered issues during fabrication and addresses geometry errors and asymmetries. In order to minimize the patch effects on the lifetime of the trapped positrons, the bonded stacks were gold electroplated to achieve a uniform gold surface. We show by simulation and analytical calculation that how positrons confinement time depends on trap imperfections. }
\end{abstract}
\pacs{85.85.+j, 52.27.Jt, 52.65.Rr}

\maketitle
\tableofcontents

\section{Introduction}
\begin{figure}[!h]
\includegraphics[width=55mm]{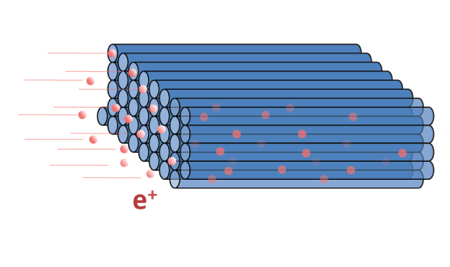}
\caption{\label{array} Schematic configuration of an array of microtraps. The image is not to scale.}
\end{figure}

The accumulation and storage of the large quantities of low-energy positrons is becoming increasingly important in different fields. Positron plasmas can be confined by static electric and magnetic fields and be in a state of thermal equilibrium for long periods of time \cite{DavidsonBook}. To accomplish the goal of energy storage, one of the fundamental limitations of conventional PM traps needs to be overcome: the required electrostatic confining potentials rise to large and impractical values as the charge stored in a PM trap is increased. A design has been proposed by one of the authors (K. G. Lynn) \cite{LynnandGreaves} in order to increase positron storage by orders of magnitude, which consists of an array of microtraps, as shown schematically in Fig. \ref{array}, with a large length to radius aspect ratio ($1000:1$) and a low confinement voltage ($10\,V$). The metallic electrodes screen the charge in each microtrap.

\begin{figure}[!h]
\includegraphics[width=80mm]{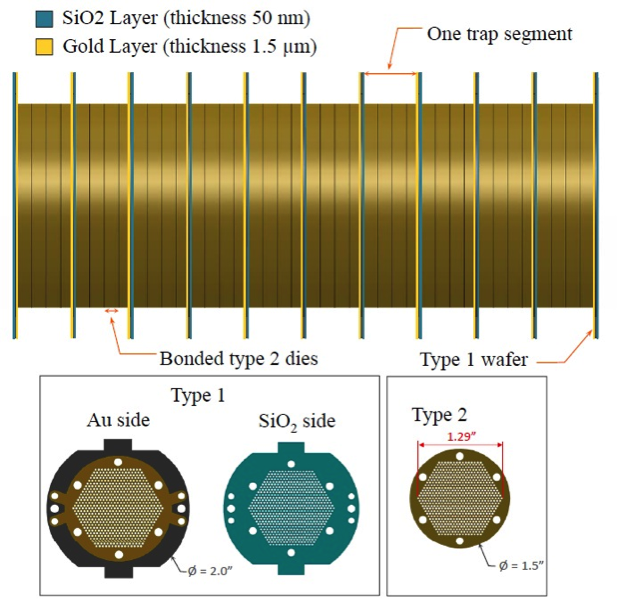}
\caption{\label{Trap} The trap configuration showing axially stacked holey dies each including $20,419$ number of $100\,\mu m$ holes. Two types of dies form the whole trap.}
\end{figure}

Since the axial confinement is assured inside the potential well between two end caps of microtrap, the plasma is lost only by radial transport across the magnetic field. Radial transport is constrained by conservation of the total canonical angular momentum of the positrons; this implies that torques from outside the plasma are required for plasma expansion and loss \cite{Simulation}. Experiments have established that the containment times depend strongly on the length of the plasma column \cite{DriscollandMalmberg} as longer plasmas experience higher levels of field and geometry asymmetries. Therefore, it is crucial to fabricate the microtraps of minimal geometry errors and asymmetries.
Analytical calculations and simulation results have shown that the trapped positron density is proportional to the inverse square of the trap radius \cite{Simulation}. As we go to smaller and smaller radii microtraps, the fabrication of the trap as well as the magnetic field alignment become more challenging. One can consider a nanotrap (as small as a cyclotron radius of positron) containing only one positron, which avoids all plasma complications and pushes the density over the Brillouin limit, and permits confinement times limited only by vacuum conditions.
The fabrication of microtrap arrays of $100\,\mu m$ diameter and $100\,mm$ length was proposed in this study. This goal can be achieved by deep etching $200$ silicon dies of $500\,\mu m$ thickness and $38\,mm$ diameter (each die contains thousands of $100\,\mu m$ holes) which were then aligned and stacked over one another to create thousands of long tubes as shown in Fig. \ref{Trap}. Computationally, more than one hundred million positrons can be trapped in one tube of this dimension ($100\,\mu m$ diameter and $100\,mm$ length) when the applied electric field to the tube ends in only $10\,V$ \cite{Simulation}. Thus, the entire trap would hold more than $10^{12}$ positrons. There were two types of dies in the trap design which create a 10-section trap ($10\,mm$ per section). The type 1 dies isolate each trap section from one another with SiO$_{2}$ as an electrical barrier enabling us to apply different voltages on trap sections which would be required to inject and accumulate positrons, while providing a physical constraint through their tab features. The Au side contains electrodes for the electrical connection. Each trap segment between two type 1 dies is fabricated by bonding type 2 dies together, which transmit the electric potential through the segment.

The ability to fabricate micro patterns and structural material for microelectromechanical systems (MEMS) using the SU-8 photolithography has been previously demonstrated \cite{Hyoung, Oh, Abgrall}. This resist has been shown to be particularly suitable for thick film applications because of its thermal stability and low optical absorption \cite{Teh}. A thick film of this resist can also be easily spin coated and exposed with conventional UV exposure systems \cite{Mata}. However, some difficulties have been reported using thick films of SU-8 including formation of edge bead \cite{Cuthbert} and air bubbles \cite{Tian} during the spin coating, as well as film adhesion issues during development \cite{Barber}.
There are several techniques for achieving deep, vertical silicon structures after fabrication of a suitable mask with the desired pattern. Cryoetching is a process relying on cooling the silicon to cryogenic temperatures, which we applied previously \cite{Ankita} for fabrication of the trap. We encountered several issues using this process such as bowing, masking, notching, and undercutting. In this study we used a Bosch process, which is a patented technique developed by Robert Bosch \cite{Franz}. There are also some other techniques based on anisotropic wet chemistry. However, they are sensitive to crystallographic orientation and they have a lower mask selectivity\footnote{Depth etched into silicon versus depth etched into mask.}.
In the Bosch process, a fluorine plasma is used to etch the silicon (etch step) after which a fluorocarbon plasma passivates the sidewalls (deposition step). Repeating these steps leads to deep and vertical etch profiles. Just as for the Cryo process, SF$_6$ gas is used in the etch step to provide a plasma of free radical fluorine. The difference between two techniques is the mechanism of sidewalls protection. In the Cryoetching process, adding oxygen to the plasma causes a layer of oxide/fluoride (SiOF) to condense on the due to the cryogenic temperatures which prevents etching of the sidewalls. While in the Bosch process C$_4$F$_8$ provides the passivation in the deposition step by breaking into longer chain radicals in the plasma and deposits as a fluorocarbon polymer. The higher process temperatures used in Bosch versus Cryoetching increases the etch rate of the mask material. Moreover, the Bosch process should be operated at higher bias voltages than the Cryoetching and it results in additional attack on the mask material and further decreases the selectivity. By tuning the parameters in each step and the time ratio of the two steps, the etch profile and the mask selectivity can be controlled.
Thin gold films are used for many applications in electronics. Following the etching, gold sputtering on the surface enables us for thermal compression bonding of wafers. However, the precise alignment of the vias during the bonding is crucial. It has been found in experiments that the confinement time of positrons in PM traps is independent of pressure when the pressure is below $10^{-7}\,Torr$, and it exhibits scaling almost as ${L_p}^{-2}$ ($L_p$ is the plasma length) \cite{Eggleston}. The anomalous loss is mainly caused by azimuthal asymmetries \cite{DriscollandMalmberg}. Hence, loses arise on experiments by trap imperfections such as misalignment of microtraps, asymmetries, and magnetic field misalignment. Simulations will help to investigate these effects and find out the amount of deviations from perfectness tolerable in our design.
Other intrinsic asymmetries, such as patch effects, are also present. The patch effects encompass various phenomena, for instance, physically imperfect surfaces (plateaus, steps, scratches, etc.), chemical impurities, and random atomic lattice orientation, which give rise to boundary regions. These all result in a variation of the local surface work function \cite{J.F.Jia} and induce local electric fields, which can influence the charged particles and might play an important role especially when the walls get very close to the particles. The effects of these potential asymmetries on the lifetime of a positron flying inside the microtrap can be very important. Gold sputtering and electroplating of the wafers inside the vias will help us to reduce these effects. Gold is favorable since it is a good electrical conductor and it is not easily oxidized.

\section{Experimental details and results}

\subsection{Photolithography}

\begin{figure}[!h]
\includegraphics[width=70mm]{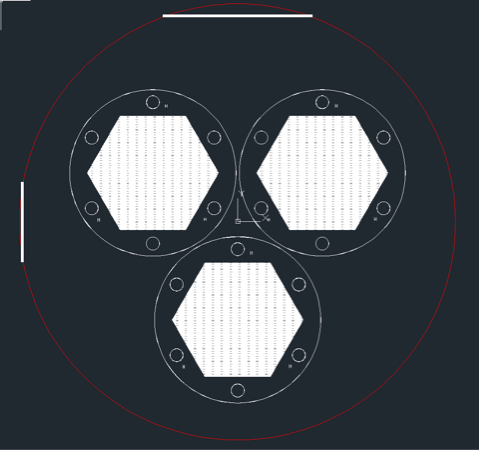}
\caption{\label{Mask} Photolithography mask.}
\end{figure}

P-type Si wafers of $100\,mm$ diameter, $540\pm5\,\mu m$ thick, $<100>$ oriented with a long primary flat and a secondary flat at a $90^{\circ}$ angle and a resistivity range of $1-20\,\Omega cm$ were used in the experiments. The trap pattern was transferred to the SU-8 photoresist onto the wafers using photolithography mask shown in Fig. \ref{Mask}.  A pattern of three dies per wafer was used, which yielded $20,419$ holes of $100\,\mu m$ diameter per each die in a hexagonal pattern with a fill factor\footnote{The fraction of total surface of holes to the trap volume.} of $0.5$. The center-to-center distance of the holes was $200\,\mu m$ and they were arranged on a $60^{\circ}$ triangular pattern. Each die was $38\,mm$ in diameter and had six $3\,mm$ holes for future alignment and bonding of the dies. Different size features reached varying depths in a given time during etching. The etch rate of trenches was adjusted to match that of holes and the size of $40\,\mu m$ for the trench width was found by experimentation in which all features are etched thorough at the same time.

To etch the features completely through the Si wafers using the Bosch process, $80\,\mu m$ thick SU-8$^{TM}$ would be required which was found by experimentation. The SU-8 2025 (from Micro-Chem Corp., Newton, MA, USA) is the least viscous photoresist of SU-8$^{TM}$ resists which can produce such a thick mask. The photolithography procedure with satisfactory results was found by doing a matrix of experiments and it is summarized in Table \ref{photolith}.

\begin{widetext}
\begin{table}[!h]
\centering
\caption{\label{photolith} Photolithography procedure of $80\,\mu m$ SU-8$^{TM}$ on $100\,mm$ diameter silicon wafers.}
\begin{tabular}{c*{14}{c}r}
\toprule
Ti adhesion layer\;\; & Dehydration bake \;\;& Spin coat (time, speed, Acc.)\;\; & Soft bake\;\; & UV Exp.\;\; & Post bake\;\; & Development\;\; \\
\hline
& & 10, 500, 100 & $5\,min$ $65^{\circ}C$ & & $5\,min$ $65^{\circ}C$ &  \\
50 nm & $5\,min$ $200^{\circ}C$ & 32, 1500, 300 & $12\,min$ $95^{\circ}C$ & $1200\,J/cm^2$ & $10\,min$ $95^{\circ}C$ & 5 min \\
& & 25 , 0, 50 & $5\,min$ $65^{\circ}C$ & & $5\,min$ $65^{\circ}C$ &  \\
\toprule
\end{tabular}
\end{table}
\end{widetext}

We found that applying a $50\,nm$ thick Ti adhesion layer via sputtering and not using the Omnicoat$^{TM}$ adhesion promoter yielded a film with a better adhesion. It was also proved that dehydrating the surface before coating was another key factor and skipping this step always caused adhesion problems. Thus, thoroughly cleaned Ti coated wafers were baked on a hotplate at $200^{\circ}C$ for $5\,min$ prior to coating. Spin coating the resist was done after allowing the substrate to reach the room temperature. Blowing nitrogen gas after installing the wafer on the spin coater and right before pouring the resist also helped to improve dehydration and to promote adhesion.
Planarization defects such as edge bead and air bubbles are more troublesome when the viscosity and thickness of the SU-8 increase \cite{Tan, Langelier}. The edge bead was removed by gently cutting it off with a clean razor blade. To prevent trapping air bobbles in the resist after spin coating due to the high viscosity of this resist, a big volume of SU-8 was poured on the center of the wafer ($\approx10\,mL$). The spinning was started when the resist approached the edge of the wafer, which normally took 1-2 seconds. Leaving the coated wafer on the spinner for 1 minute helped to relax the SU-8. Sometimes, we ended up with small bobbles which went away during first step of the soft baking process ($65^{\circ}C$). Almost one out of ten wafers was not bublle-free after soft baking using this technique. Spin coating produced fairly uniform films with $2\,\mu m$ deviation. However, spin chuck vacuum was causing a circular mark at the center due to the deforming of the substrate. The chuck mark was avoided to some extent by removing the O-ring on the spin chuck.
During the prebake step, the film temperature was brought above the glass transition temperature of SU-8 on the hotplate for enough time to remove the solvents and also relieve stresses incurred by spin coating. Prebaking on the hotplate was preferred because it reduces the chance of trapped solvents inside the film. Underbaked resists resulted in partial or complete delamination of the mask during development and also caused to patterns with lower resolutions. When the wafer stuck to the mask after the exposure it was a sure sign of underbaking. On the other hand, overbaking the resist made it nearly impossible to be developed. Two temperature ramps were used to reduce stress in the resist as shown in Table \ref{photolith}. Improper ramp time and temperature resulted in delamination or micro-cracking during development.
After a $10\,min$ relaxation step with the wafer on a level, nonmetal surface, the glass photolithography mask containing the pattern shown in Fig. \ref{Mask} was placed over the wafer and was exposed with UV light at $1200\,J/cm^2$. After exposure, the wafer was transferred to the hotplates for a post-exposure bake. Then it was left for another $10\,min$ to cool down to room temperature and relax.
Development of the SU-8 film, which is a chemical removal process, was done using SU-8 developer$^{TM}$ (from Micro-Chem Corp., Newton, MA, USA). All developments were performed with gentle agitation of solution. The main problem during the development was the mask delamination. It has been known that the adhesion is due to process parameters such as the soft bake and the post-exposure bake times and temperatures, exposure level, and development times \cite{Barber}. The adhesion quality of SU-8 mask was traced carefully during the development as a matter of delamination time and extent. Observation during matrix experiments after each step of process were also recorded which helped to determine the conditions that would lead to good adhesion without any delamination. We did not have any success without using a sputtered Ti adhesion layer. Even applying HMDS (hexamethyl disilane) and/or removing the Si oxide layer to promote adhesion were fruitless. The large difference of thermal expansion coefficient between Si ($2.6\,ppm/K$) and SU-8 ($50\,ppm/K$) causes high values of internal stress in the mask as the resist is heated and cooled several times and it becomes more brittle as the solvent level decreases. Applying a Ti seed layer substantially decreased this stress (thermal expansion coefficient of Ti is $8.3\,ppm/K$) and along with optimized processing parameters led to good adhesion of the SU-8 film.

\subsection{Deep reactive ion etching}

\begin{figure}[!h]
\centering
\begin{subfigure}{.5\textwidth}
  \centering
  \includegraphics[height=30mm]{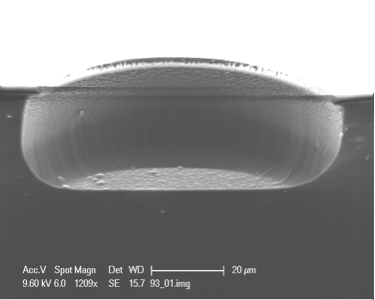}
  \caption{isotropic etch due to the improper SF$_6$ and C$_4$F$_8$ flows ($85\,sccm$ and $60\,sccm$)}
\end{subfigure}
\begin{subfigure}{.5\textwidth}
  \centering
  \includegraphics[height=30mm]{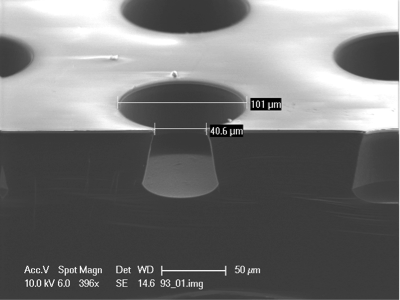}
  \includegraphics[height=30mm]{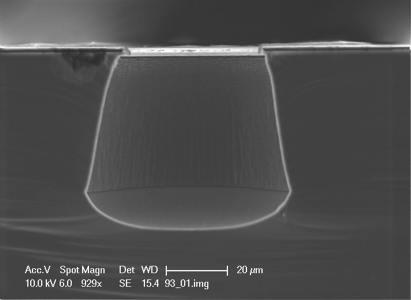}
  \caption{Negative tapering (deviation from $90^{\circ}$ angle) due to the improper ratio of etch and deposition time steps}
\end{subfigure}
\begin{subfigure}{.5\textwidth}
  \centering
  \includegraphics[height=30mm]{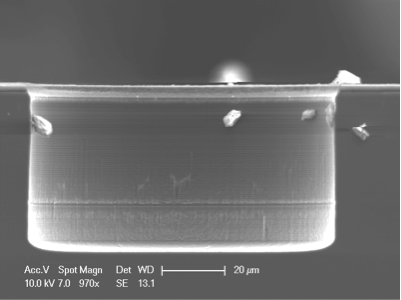}
  \includegraphics[height=30mm]{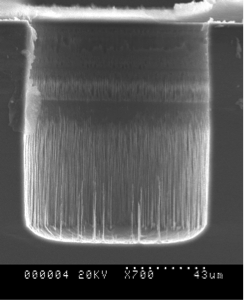}
  \caption{Micromasking and rough walls at deeper sections due to using high pressure ($30\,mTorr$).}
\end{subfigure}
\caption{Examples of unsuccessful etchings.}
\label{badetch}
\end{figure}

The exposed Si from the photolithography process was then etched by DRIE. An Oxford Plasmalab$^{TM}$ 100 machine (Oxford Instruments plc, Abingdon, Oxfordshire OX13 5QX United Kingdom) was used for this process. The machine has an RF plasma power of up to $300\,W$ at $13.56\,MHz$, ICP power up to $3,000\,W$ at $13.56\,MHz$, and a substrate temperature range from $-150$ to $+400^{\circ}C$.
Achieving a good process requires an understanding of the mechanisms occurring. Initially, the ICP power ranging from 450 to $1600\,W$ and an RF power ranging of 10 to $50\,W$ was studied. These power values were limited to avoid burn the resist. The SF$_6$ and C$_4$F$_8$ flow was adjusted so that the profile was anisotropic. The temperature of the substrate holder was maintained at $15^{\circ}C$ and the helium backside flow ensured a good thermal conduction to the substrate. Adjusting the process was done by changing the ratio of each step in order to achieve a vertical profile by balancing between the polymer deposition and etching without changing the ions energy. Pressure is one the key parameters in this technique, which substantially affects the plasma properties and so the surface chemistry. Figure \ref{badetch} shows examples of unsuccessful etches due to the improper SF$_6$ and C$_4$F$_8$ flows, pressure, and ratio of etch and deposition time steps. Above a certain flow the etch rate dropped and the profile became anisotropic.

\begin{figure}[!h]
\centering
\begin{subfigure}{.5\textwidth}
  \centering
  \includegraphics[width=55mm]{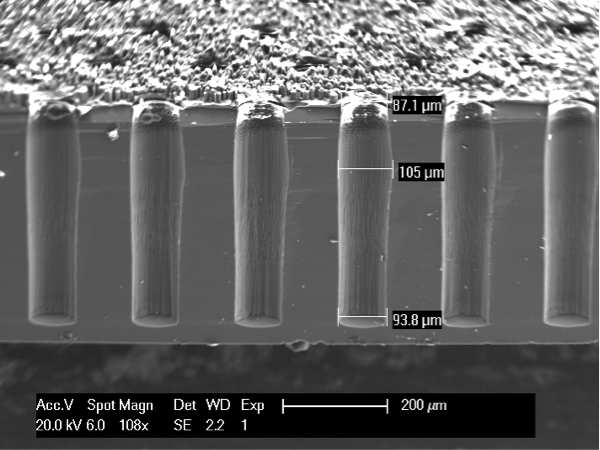}
  \caption{Bowing effect of $20\%$ at the top section of the holes. This sample is not etched completely through the thickness. }
\end{subfigure}
\begin{subfigure}{.5\textwidth}
  \centering
  \includegraphics[width=55mm]{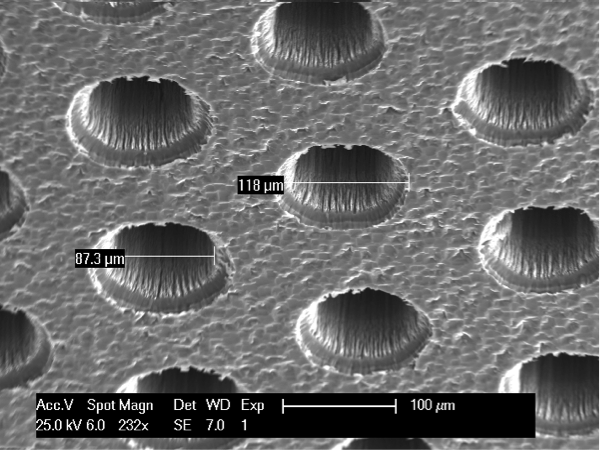}
  \caption{Backside of the sample which is etched thoroughly showed a back-notching due to the excessive etching. Wafer backside is not polished.}
\end{subfigure}
\caption{The etch profile using the optimized Bosch process recipe in Table \ref{Bosch}. }
\label{goodetch}
\end{figure}

\begin{figure}[!h]
\includegraphics[width=55mm]{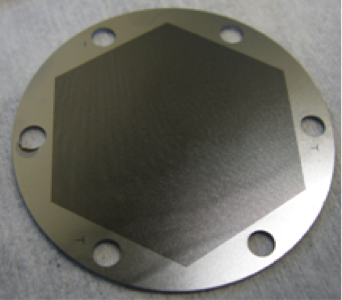}
\caption{\label{die} One die etched completely through. All the features were resolved.}
\end{figure}

Running a series tests with different parameters led us to the recipe shown in Table \ref{Bosch} while the corresponding etch profile is shown in Fig. \ref{goodetch}. Lower pressures allowed better controlling of the profile since it reduces the ion scattering by collisions. Decreasing the pressure also prevents the gas interference \cite{Douglas}. So we were able to etch through the whole thickness without micromasking of the walls at the bottom. A semiconductor tape on the backside of the wafers was added and we were able to get all features resolved in the wafer as shown in Fig. \ref{die} while simultaneously separating the dies from wafer. The tape also aided in keeping the wafer and dies together while it was removed from the DRIE machine. Increasing the SF$_6$ gas flow along with lowering the pressure, which decreases the radicals density, caused to increase the etch rate from $1.2$ to $2.1\,\mu m/min$. The mask selectivity of $7:1$ was obtained in this run. Trenches with different widths were processed and it revealed that the trench width of $40\,\mu m$ had the same etch rate as the $100\,\mu m$ holes. This is important as we needed the all features to be etched completely through the wafer at the same time.

\begin{widetext}
\begin{table}[!h]
\centering
\caption{\label{Bosch} Optimized Bosch process recipe. The corresponding etch profile is shown in Fig. \ref{goodetch} exhibits a $20\%$ bowing defect.}
\begin{tabular}{c*{14}{c}r}
\toprule
& SF$_6$ flow  \;\;& C$_4$F$_8$ flow \;\; & Time step \;\; & Pressure\;\; & Backside He pressure\;\; & ICP/RF power \;\; \\
& ($sccm$)  \;\;& ($sccm$) \;\; & ($s$) \;\; & ($mTorr$)\;\; & ($Torr$)\;\; & ($W$) \;\; \\
\hline
Etch & 100 & 1 & 28 & 26 & 7 & 1500/25 \\
Deposition & 1 & 100 & 14 & 20 & 7 & 1600/10 \\
\toprule
\end{tabular}
\end{table}
\end{widetext}

\begin{figure}[!h]
\includegraphics[width=55mm]{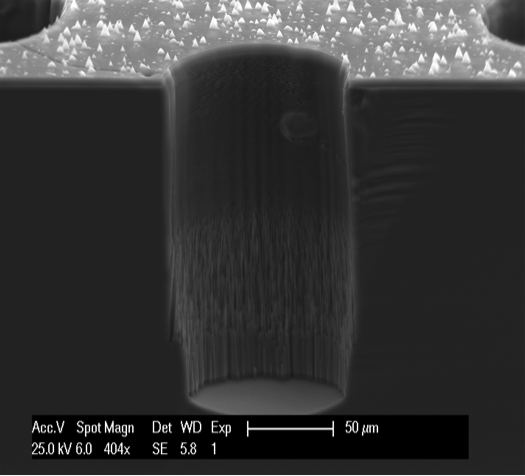}
\includegraphics[width=55mm]{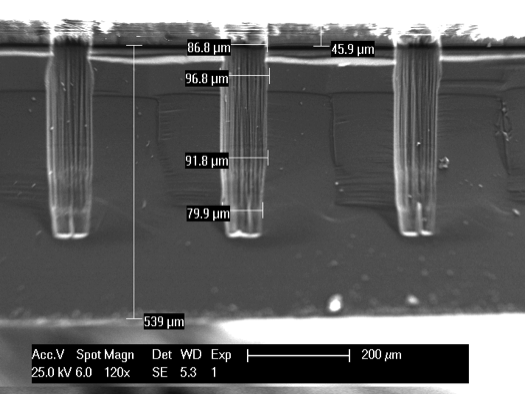}
\caption{\label{shortpulse} The etch profile with decreased etch and deposition time steps ($12\,s$ and $6\,s$) while other parameters kept constant as in Table \ref{Bosch}. }
\end{figure}

The minimum length of etch and deposition steps were restricted by response speed of the pressure and mass flow controllers and also the error lapse time in the software. This time was 14 seconds for Oxford Plasmalab$^{TM}$ 100 machine which was found by experimentation. In the following experiment we decreased the cycle times to $12\,s$ and $6\,s$ for etch and deposition steps respectively, where the ratio of two were equal to previous cases. Figure \ref{shortpulse} exhibits the profiles obtained with this process. While less bowing defect was experienced using decreased time steps, strong striations were seen at the walls and the profile showed micromasking defects at deep sections. As the depth of the etch increased, the attack on the sidewalls due to ion bombardment became less and so resulted in polymer buildup and micromasking at the bottom of the holes. The mask selectivity was also improved in this experiment from $7:1$ to $33:1$ which would help us by utilizing a thinner resist and improve the bowing defect even further by avoiding ions deflection charged by hitting the resist walls. The mask selectivity was almost insensitive to the nature of the resist (as we also tried other resists such as AZ photoresists) and even hard baking of the resist prior to etching did not improve it. Sidewall striation observed in the results was a sign of overpassivated sidewalls of the holes. Increasing the RF power and so the bias voltage in the etching step would improve the passivation layer removal from the sidewalls and prevent micromasking but it would also increase ions attack to the walls and consequently worsen the bowing defect or result in a positive taper.

In the recent experiment, the pulses were not long enough to stabilize the plasma and obtain a reasonable duty cycle. Our Oxford Plasmalab$^{TM}$ 100 machine was originally designed for the Cyroetching process and lack some of the equipment used for the Bosch technique. Firstly, this process needs fast response mass controllers which exceed SEMI standard E17-91. The response time of these controllers should be below one second as the flow reaches within $2\%$ of the set point. Also, the time delay between opening these controllers and reaching the gas to the chamber are minimized in these systems. These allow having short etch and deposition time steps which was not doable in our system.
Due to the fact that we were not able to eliminate the bowing defect completely and also that we would need a high rate of production to fabricate all 200 silicon dies needed to complete the trap, we got the entire samples with the same pattern from RTI International Institute and we continued fabrication of the trap by gold-coating and bonding the dies at our research center at Washington State University (WSU). All of the dies were inspected thoroughly and those which had more than 10 plugged holes in total or had excessive back notching were rejected.

\subsection{Gold sputtering}

\begin{figure}[!h]
\includegraphics[width=55mm]{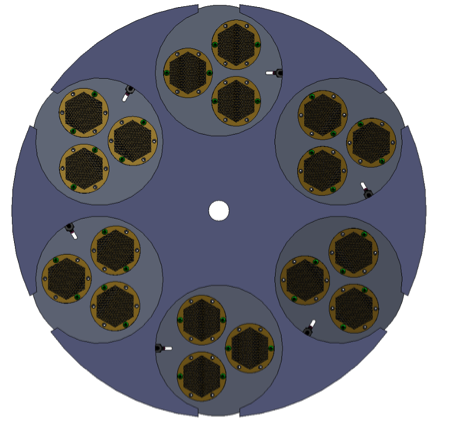}
\includegraphics[width=55mm]{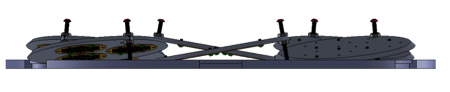}
\caption{\label{sputtering} Bottom view (top image) and side view (bottom image) of the sputtering configuration illustrating the dies rotating off-axis of gold target while it was kept with an angle from horizontal state. }
\end{figure}

In order to bond the dies together and fabricate the segments shown in Fig. \ref{Trap}, we sputtered a thin film of gold on the dies surfaces ($1.5\,\mu m$). The native oxide layer of silicon was stripped first with dipping into the BOE (Buffered oxide etch of NH$_4$-HF $10:1$) solution for $10\,s$. A TiW layer of $40\,nm$ thick was sputtered to get good adhesion of the gold on the samples which was qualitatively checked by the typical tape test. It was also crucial to have gold coated on the inside of the holes to obtain a uniform potential throughout the holes. A sample die was gold coated flat and parallel to the sputtering target. However, the result showed a resistivity of $4\,kOhm$  through the holes due to the absence of gold inside the holes. The solution resided on sputtering both side of the dies with an angle ($tan^{-1}(D/L)\approx 11^{\circ}$, where $D$ is the diameter and $L$ is the length of the holes) from horizontal state while it was rotating off-axis of target as illustrated in Fig. \ref{sputtering}. The flipping axis to the second side of the die was also important to get all the inside surfaces coated.

\begin{figure}[!h]
\centering
\begin{subfigure}{.5\textwidth}
  \centering
  \includegraphics[width=55mm]{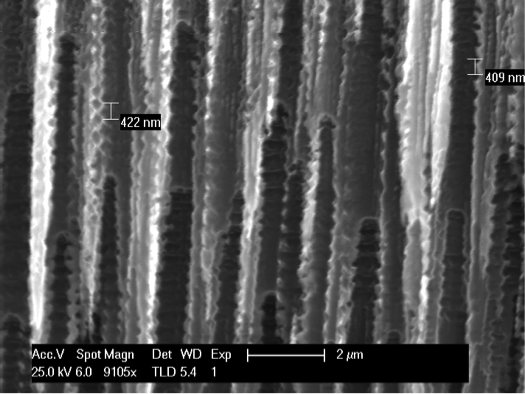}
  \caption{The roughness of bare silicon inside the holes of not gold coated die due to the Bosch etching process with scalloping size of more than $400\,nm$.}
\end{subfigure}
\begin{subfigure}{.5\textwidth}
  \centering
  \includegraphics[width=55mm]{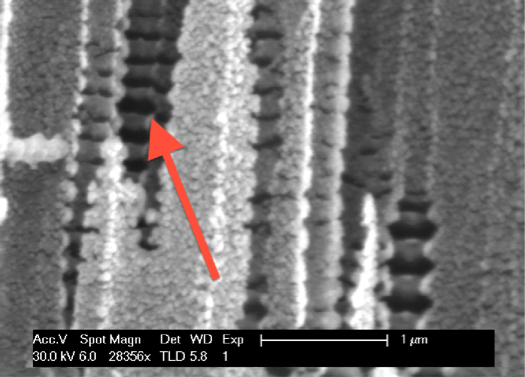}
  \caption{The gold grains are visible at the surface inside the holes of gold coated die. The arrow shows a black micro-cave that was not necessarily coated with gold.}
\end{subfigure}
\caption{SEM image of a $100\,\mu m$ hole before and after gold sputtering. }
\label{sputtered}
\end{figure}

The electrical tests using an ohmmeter proved continuity through the holes but this method did not demonstrate anything about the uniformity of the coating. We also had concerns about electric potential variations from hole-to-hole. Thus, the sample die was cleaved after sputtering both sides. The broken area of the die exposed several of the holes. By applying a $9.4\,VDC$ bias to the die and using a voltmeter with a very fine wire (48 AWG), we were able to probe the exposed holes. Each of the holes confirmed the $9.4\,VDC$ bias. It was also noted that probing the bare silicon between the holes gave a much lower voltage reading ($\approx 3\,VDC$). To further check the consistency between holes, the 48 AWG wire was used to probe additional holes in the die. A $2\,mm$ length was stripped at the end of the wire, the wire was pushed all the way through the hole, and the $2\,mm$ tip was then bent and the wire was slowly pulled back through the hole. This provided good contact to the hole surface while observing the voltage as the wire was pulled through the hole. Several holes were checked in this manner across the die and all read a constant $9.4\,VDC$ while the tip of the wire was inside the hole.
The SEM image in Fig. \ref{sputtered} shows inside the holes before and after gold sputtering. The scalloping size of the walls due to the Bosch process was measured about $400\,nm$. After gold sputtering the sample with the described technique, the gold grains were detected all over the surface except in some micro-caves with dimension of $500\,nm$ at maximum. These caves were mostly found near the middle depth of the holes where the gold ions had difficulty reaching.

\subsection{Bonding}

\begin{figure}[!h]
\includegraphics[width=80mm]{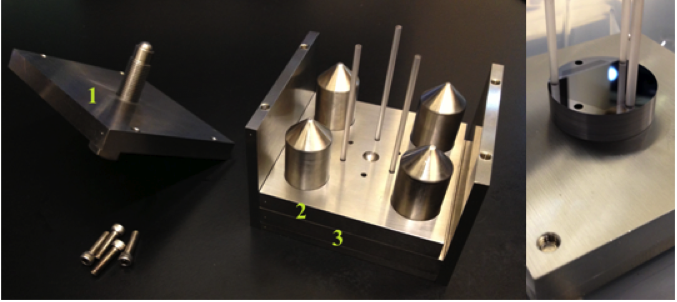}
\caption{\label{jig} The aligner-bonder jig used in the experiments. The dies were stacked through the sapphire rods in between plate (2) and (3). The screw in the plate (1) pressed plate (2) against plate (3). }
\end{figure}

After gold sputtering the dies, we used gold thermocompression bonding to bond the etched dies. Initially, the gold coated dies were cleaned. An alignment and bonding jig made of Invar36 shown in Fig.\ref{jig} was precision-machined and used for the bonding. The top plate (1) contained a screw that presses plate (2) against plate (3). The dies were threaded onto the sapphire rods in between plate (2) and (3). The sapphire rods used for bonding were $3.004\pm 0.001\,mm$ diameter and $50\,mm$ long with a straightness of $6\,\mu m$ over the entire length. There were two kinds of $3\,mm$ holes on the Si dies. Three out of six holes, named Tight-tolerance holes, had diameter of $3.0096\pm 0.0005\,mm$, were utilized with the sapphire rods when bonding took place. The other three holes, named as Regular holes, had diameter of $3.0299\pm0.0005\,mm$, and used for handling and coarse alignment of dies and bonded segments.

\begin{figure}[!h]
\includegraphics[width=70mm]{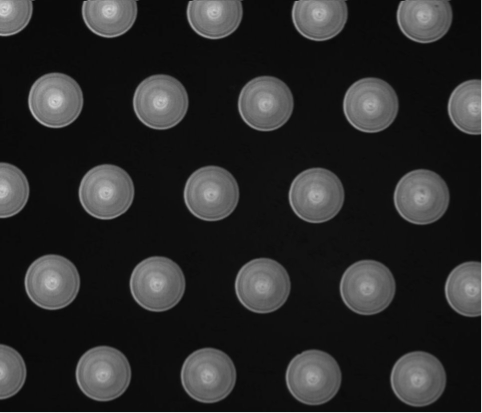}
\caption{\label{IR} IR transmission microscope image from the stack of thirteen dies, which were gold coated, aligned and then bonded. }
\end{figure}

\begin{figure}[!h]
\includegraphics[width=42mm]{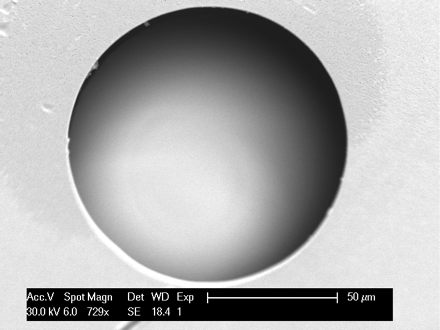}
\includegraphics[width=42mm]{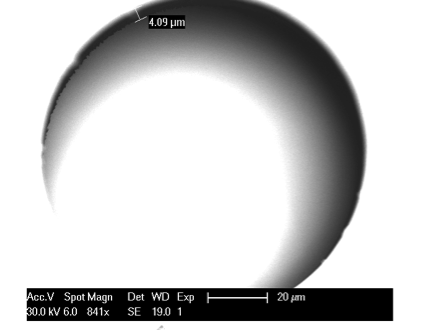}
\includegraphics[width=42mm]{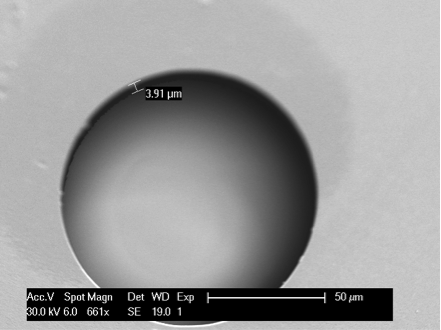}
\includegraphics[width=42mm]{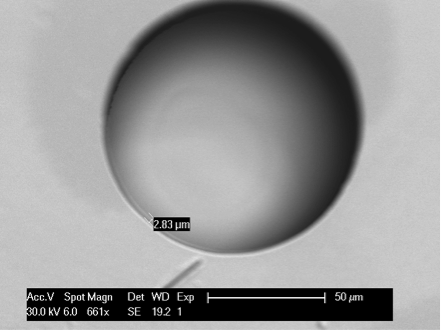}
\caption{\label{bondedSEM} SEM images from the stack of thirteen dies, which were gold coated, aligned and then bonded. }
\end{figure}

The entire jig along with the dies were loaded into a furnace (Barnstead Thermolyne 1400) that was stable at the bonding temperature. The jig (with the dies and sapphire rods) was baked for $60\,min$ to allow coming to equilibrium before adding pressure to the dies. Several bonding attempts were made at the temperatures ranging from 200 to $300^{\circ}C$ and pressures from 16 to $1230\,psi$ and bonding time from 20 to $60\,min$. Many authors report either much higher bonding temperatures ($\approx 400^{\circ}$C) or much higher pressures ($\approx 1,800\,psi$) \cite{Jing, Wang}. Successful bonding in our jig occurred at temperature of $250^{\circ}C$ and pressure of $1,230\,psi$ ($9.7\,kN$) for $60\,min$ and proved to be a reliable and reproducible process. One experiment yielded a bonded stack of thirteen etched dies. Analysis was needed to evaluate the quality of the bond and alignment precision. A rudimentary drop test from a $5\,cm$ height was performed and successfully demonstrated reliable bond strength. An IR transmission microscope image from an area of the stack is shown in Fig. \ref{IR}, representing a very successful alignment. The SEM images were taken from an example hole, though it was not possible to focus on all 13 dies. The misalignment of the dies was measured at about $4\,\mu m$, as shown in the Fig. \ref{bondedSEM}.

\subsection{Gold electroplating}

The bonded stacks were then gold electroplated in order to fill the micro-caves which might not be coated during gold sputtering shown in Fig. \ref{sputtered}. The '24k Pure Gold' gold plating solution (from Gold Plating Services company, Kaysville, Utah) was used in experiments which is designed to produce a gold electrodeposit with higher purity than $99.9\%$. The samples surfaces were cleaned and activated with HCl acid for $5\,min$ prior to electroplating. A Platinized Titanium plates used as anodes (also from Gold Plating Services). A series of experiments were done to find the optimum parameters, which gave a good throwing power through the holes, and the current density found to be the most important parameter. A moderate agitation also helped circulating of the solution inside the holes. The successful electroplating of $2\,\mu m$ thick gold is summarized in Table \ref{plating}.

\begin{table}[!h]
\centering
\caption{\label{plating} Parameters used in electroplating of $2\,\mu m$ thick gold.}
\begin{tabular}{c*{14}{c}r}
\toprule
Temperature\; & Anode to Cathode  \; & Current Density \; & Time \\
($^{\circ}C$) \;& Ratio \; & ($mA/in^2$) \; & ($min$) \\
\hline
$55\pm1$	&  $3.5:1$	&20	&10 \\
\toprule
\end{tabular}
\end{table}

\begin{figure}[!h]
\includegraphics[width=70mm]{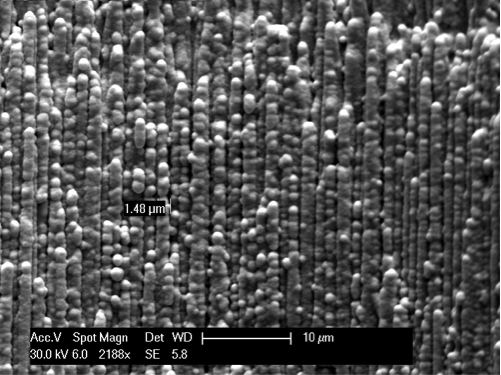}
\caption{\label{platedSEM} SEM image of electroplated holes exhibiting uniform plating. }
\end{figure}

\begin{figure}[!h]
\includegraphics[height=30mm]{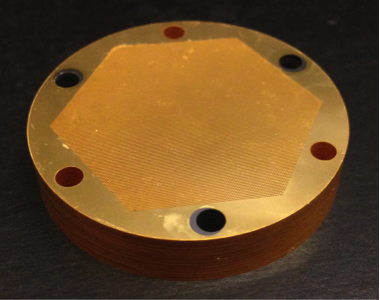}
\includegraphics[height=30mm]{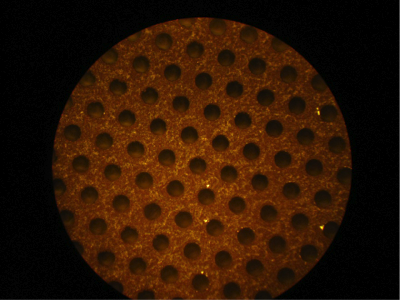}
\caption{\label{plated} The electroplated stack of 13 dies. }
\end{figure}

One single die was electroplated with the described process. The quality of the plated layer inside the holes was studied with SEM after cleaving the die, shown in Fig. \ref{platedSEM}. The results showed a uniform gold layer throughout the holes and convinced us that we could also have the bonded stacks electroplated with the same parameters and enough gold would be plated even inside the longer holes. Figure \ref{plated} shows the bonded stack of 13 dies, which was also electroplated. Due to the grain structure of this type of gold deposit the reflective qualities of the surface became matt.

\section{Effects of trap imperfections on confinement time of positrons}

\subsection{Patch effects}

If the entire surface inside the tubes is coated with gold with no roughness, the effect of potential asymmetries on the lifetime of a positron flying inside the microtrap can be estimated. With a work function variation ($\Delta\phi$) of less than $1\,mV$ for an evaporated gold surface, and estimation of the RMS potential variation along the axis of a cylindrical electrode, $RMS\,\Phi$, as \cite{Camp}
\begin{equation}\label{RMSdeltaV}
RMS\,\Phi=\frac{0.6\,\Delta\phi \,l_{c}}{R_w},
\end{equation}
it is calculated that $RMS\,\Phi<10^{-6}\,V$, when $R_w = 50\,\mu m$ and patch length, $l_c$, is comparable to the grain size of sputtered gold onto a silicon made microtrap,  $0.1\,\mu m$.

The perpendicular drift velocity of a positron due to the patch field can be assumed as
\begin{equation}\label{patch1}
V_{\bot}=\frac{E}{B},
\end{equation}
and also the movement as
\begin{equation}\label{patch2}
\Delta x=V_{\bot} \frac{l_{c}}{V_0},
\end{equation}
where $V_0$ is the velocity by which the positron passes over the patch length (almost equal to the total velocity in a high magnetic field). The movement due to $N$ equal patches can be written then as
\begin{equation}\label{patch3}
\overline{X}^2=N{(\Delta x)}^2.
\end{equation}
Since $Nl_c=V_0t$, by substituting $N$ and $\Delta x$ in Eq. (\ref{patch3}) we obtain $t$ as
\begin{equation}\label{patch4}
t=\frac{B^2 {\overline{X}}^2 V_0}{E^2 l_c}.
\end{equation}
One calculates  $t\approx4000\,s$, the time for the positron to get from the microtrap axis to the gold coated wall when  $\overline{X}=R_w=50\,\mu m$ is the microtrap radius,  $B=7\,T$,  $V_0= 1.32\times 10^{6}\,ms^{-1}$ for a $5\,eV$ positron,  $l_c=0.1\,\mu m$, and  $E=(1\,mV)/(50\,\mu m)=20\,Vm^{-1}$. The patch effects can be more problematic when there are areas of bare silicon, which result in more variation of the local surface work function. Thus we subjected the dies to gold electroplating after bonding in order to fill the micro-caves with gold.

\subsection{Dies and magnetic field misalignments}

\begin{figure}[!h]
\includegraphics[width=70mm]{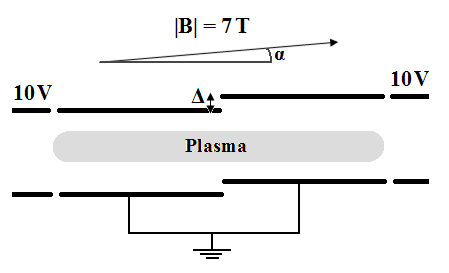}
\caption{\label{model} The schematic geometry of a microtrap and a plasma. The image is not to scale.}
\end{figure}

Magneto- and electrostatic errors which arise from misalignment of the trap cylinders or misalignment of the magnetic field direction with the microtrap longitudinal axis could be very problematic in microtraps. It has been reported \cite{DriscollandMalmberg} that with improvements in conventional PM traps fabrication and less sectors misalignment ($0.1\%$), the trapped particles survived longer in experiments. Simulation conducted here helped us to better understand the importance of these parameters.
Modeling simulations were carried out with WARP, a code used extensively in plasma physics \cite{WARP}. In WARP, the particle-in-cell (PIC) method is employed. A discrete number of real particles Ð positrons in this case Ð are combined to so called macro-particles. The Lorentz equation of motion is employed to advance macro-particles in time. Following each time step, the charge density is calculated via a linear interpolation of the macro-particles position onto a mesh. By solving PoissonÕs equation, the electrostatic potential is then calculated from the charge density.  Since the rotational symmetry of the microtrap is broken in our cases, the three dimensional version of WARP is used for simulation. It uses a $xyz$ field solver with constant potential boundary conditions (i.e. Dirichlet conditions) at the electrode walls.
A schematic of a $50\,\mu m$ radius microtrap modeled in our simulations is shown in Fig. \ref{model}. Modeling parameters are listed in Table \ref{parameters}.  The microtrap is composed of a central perfectly conducting, grounded tube and two end electrodes. The microtrap is immersed in a uniform, constant magnetic field of $7\,T$. The end electrodes potential are constant at $10\,V$. The startup parameters of the plasma played a vital role in reducing the computational effort. The major simulation parameters such as time step and mesh size were chosen carefully. Discordant values of them cause a large numerical instability. The choice of values for the parameters with the largest influence on numerical noise is discussed in another study \cite{Simulation}. The initial uniform plasma radius is $1/\sqrt{3}$ of the microtrap radius and its positron density is equal to $4.8\times 10^{11}\,cm^{-3}$. With this density the space charge potential at the plasma axis is equal to $3.75\,V$ which is confined axially by $10\,V$ end electrodes.

\begin{table}[!h]
\centering
\caption{\label{parameters}The modeling parameters of the simulation.}
\begin{tabular}{c*{14}{c}r}
\toprule
&\multicolumn{7}{c}{Modeling parameters} \\
\cline{2-8}
Case &  $\Delta t \footnote[1]{Time step} $ & $\Delta R\footnote[2]{Mesh size}  $ & $PW\footnote[3]{Positron weight: the number of real particles that each simulation macro-particle represents.} $ & $T_0 $ & $\Delta$  & $\alpha$ &  $L_g\footnote[4]{Overal length of the central tube} $  \\
No. &$(ps)$ & $(\mu m)$ & & $(eV)$ &  $(\mu m)$  & $(^{\circ})$ & $(mm)$ \\
\hline
D1	&2.5&	3.35&	5&	0.5&	0&	0&	1 \\
D2	&2.5	&       3.35&	5&	0.5&	2&	0&	1 \\
D3	&2.5	&       3.35&	5&	0.5&	5&	0&	1 \\
B1	&2.5&	3.35&	5&	0.5&	0&	0.01&	5 \\
B2	&2.5&	3.35&	5&	0.5&	0&	0.05&	5 \\
B3	&2.5&	3.35&	5&	0.5&	0&	0.1&	5 \\
B4	&2.5&	3.35&	5&	0.5&	0&	1&	0.5 \\
B5	&2.5&	3.35&	5&	0.5&	0&	2&	0.5\\
\toprule
\end{tabular}
\end{table}

\begin{figure}[!h]
\includegraphics[width=70mm]{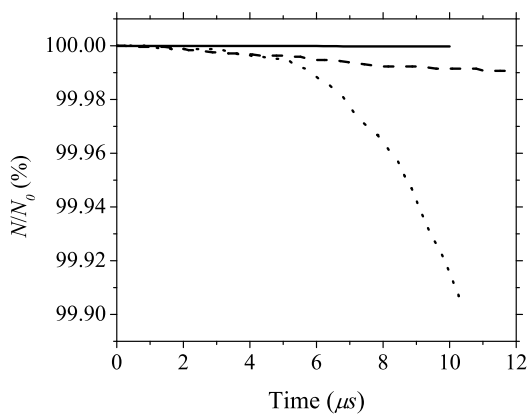}
\caption{\label{diemisalignmentloss} The time histories of the number of trapped positrons in the cases D1, solid line, D2, dashed line, and D3, dot line.}
\end{figure}

The central tube includes two $0.5\,mm$ long sectors misaligned to each other at the cases D2 and D3 while the magnetic field is perfectly axial ($\alpha = 0$). The misalignment, $\Delta$, is equal to $2\,\mu m$ and $5\,\mu m$, respectively. Figure \ref{diemisalignmentloss} compares the histograms of trapped particles in these cases to the symmetrical microtrap in the case D1 with $\Delta=  0$ and $\alpha = 0$. Higher misalignment caused increasing the loss rate of positrons. When the misalignment is $5\,\mu m$, the loss rate accelerated in time. Almost linear loss rate was experienced for the $2\,\mu m$ misalignment, suggesting that the only one percent of positrons would be lost after $1\,s$.

\begin{figure}[!h]
\includegraphics[width=85mm]{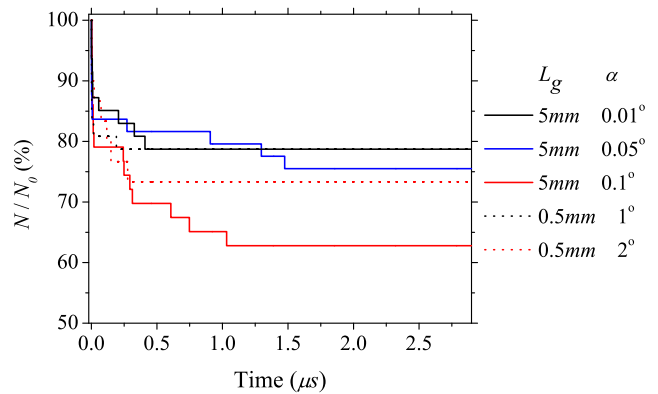}
\caption{\label{Loss}The fraction of trapped positrons as the WARP simulation proceeds for different values of $L_g$ and $\alpha$.}
\end{figure}

\begin{figure}[!h]
\includegraphics[width=75mm]{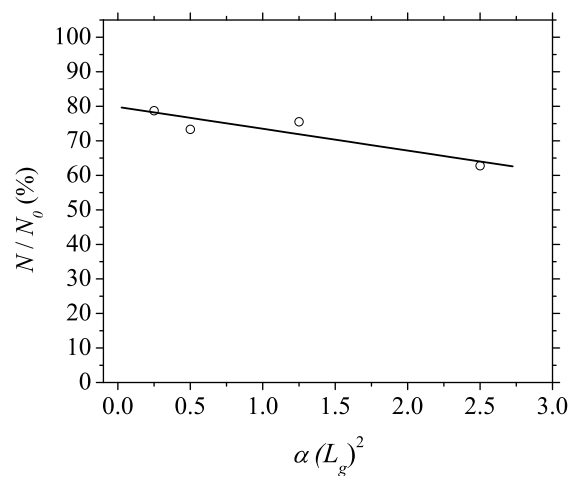}
\caption{\label{trend}The fraction of trapped positrons versus the value of $\alpha (L_g)^2$. The data is fitted with the linear function, $y=ax+b$, when $a=-6.32\pm1.53$ and $b=79.81\pm1.96$.}
\end{figure}

In the cases B1 through B5, the main tube is a perfect cylinder with $\Delta=  0$ . The simulation used two different lengths of the tube, $L_g$, equal to $0.5\,mm$ and $5\,mm$, which are consistent with underway experiments in WSU, equivalent to 1 and 10 dies of silicon holes with $500\,\mu m$ lengths, respectively. Fig. \ref{Loss} shows the loss due to the magnetic field misalignments in different simulation runs. The percentage of trapped particles obtained is plotted versus the value of $\alpha (L_g)^2$  in Fig. \ref{trend}. The linear fit of data suggests that for the microtrap of $10\,cm$ length and $100\,\mu m$ diameter the magnetic field should be aligned with an accuracy of ${2\times10^{-4}}^\circ$ to get more than $60\%$ of initial particles trapped for long times.

\section{Conclusion}
The SU-8 photolithography was studied to obtain $80\,\mu m$ uniform resist on $100\,mm$ Si wafers. We have also studied Bosch process for deep reactive ion etching of high aspect ration holes. The trench width were optimized in order to resolve the entire features in the same time and avoid back notching at the hole bottoms. The Bosch etching technique proved to be successful process for anisotropic etching although we were not be able to eliminate the bowing effect completely due to the equipment limitation. The mask selectivity and the bowing defect were improved by reducing the etch and deposition step times (while the ratio of two steps kept constant) although strong striation was experienced since the pulses were not long enough to stabilize the plasma and obtain a reasonable duty cycle.
We were able to gold sputter the entire surface inside the holes by sputtering the rotating dies on angle. However, some micro-caves were detected which were covered with gold during the gold electroplating of bonded stacks. We estimated the patch effects addressed in this study on the confinement times of the positrons and so the holes were gold electroplated in order to fill the micro-caves. More realistic effect of these stray electric fields on the lifetime of a confined particle ensemble in the plasma regime could be a subject of simulation while it is not expected to be a dominant factor since variations in the tube diameter, which are about $\approx 2\,\mu m$ with the current fabrication process, play a bigger role than that calculated from Eqs. \ref{RMSdeltaV} and \ref{patch4}. The work function of the electroplated surface will be measured by Kelvin probe technique. Smoother gold thin films are reported to be formed by thermal annealing at temperature of $160^{\circ}C$ or above in the UHV chamber \cite{Nan}.
Aligning and bonding of the dies were done using a precision-machined jig. Bonding proved to be successful while a misalignment of $4\,\mu m$ was achieved. Particle-in-cell WARP simulations showed that confinement of pure positron plasma columns in microtraps depends strongly on the alignment of the dies and magnetic field. Dies misalignments higher than $2\,\mu m$ affect largely on the storage capacity of the microtraps and cause accelerated losses due to the broken symmetry. Therefore, we will try to improve the alignment further during bonding. We also need to align the magnetic field carefully with microtraps axis. For the small values of field misalignment, the percentage of the trapped particles scaled with the value of $\alpha^{-1} (L_g)^{-2}$, where $L_g$ is the length of the trap in millimeters and $\alpha$ is the misalignment angle in degrees. Further experimental results using the fabricated array of microtraps to store charged particles will be published once the desired goals are achieved.

\begin{acknowledgments}
The authors are grateful to Randall Svancara for his helps to the simulations on HPC. We are also thankful to Dr. David Grote at LLNL for assistance and useful discussions regarding the WARP simulations. We would also like to thank program managers Dr. William Beck and Dr. Parvez Uppal of the Army Research Laboratory who provide funding under contract $W9113M-09-C-0075$, Positron Storage for Space and Missile Defense Applications, and program manager Dr. Scott Coombe of the Office of Naval Research who provide funding under award $\#N00014-10-1-0543$, Micro- and Nano-Traps to Store Large Numbers of Positron Particles at Very Large Densities.
\end{acknowledgments}

\bibliographystyle{unsrt}
\bibliography{myreference}

\end{document}